\documentclass[12pt]{article}
 \usepackage{epsfig}
 \def\be{\begin{equation}}
 \def\ee{\end{equation}}
 \def\bea{\begin{eqnarray}}
 \def\eea{\end{eqnarray}}
 \usepackage{graphicx}

 \catcode`\@=11
 \def\lsim{\mathrel{\mathpalette\@versim<}}
 \def\gsim{\mathrel{\mathpalette\@versim>}}
 \def\@versim#1#2{\vcenter{\offinterlineskip
 \ialign{$\m@th#1\hfil##\hfil$\crcr#2\crcr\sim\crcr } }}
 \catcode`\@=12

 \catcode`@=12
\begin{document}
 \thispagestyle{empty}
 \begin{flushright}
 UCRHEP-T600\\
 Dec 2019\
 \end{flushright}
 \vspace{0.6in}
 \begin{center}
 {\LARGE \bf Leptonic Source of Dark Matter and\\ 
Radiative Majorana or Dirac Neutrino Mass\\}
 \vspace{1.2in}
 {\bf Ernest Ma\\}
 \vspace{0.2in}
{\sl Physics and Astronomy Department,\\ 
University of California, Riverside, California 92521, USA\\}
\end{center}
 \vspace{1.2in}

\begin{abstract}\
The notion of U(1) lepton number (which may only be softly broken) is applied 
to models of dark matter which interacts with leptons.  Previous scotogenic 
models of Majorana or Dirac neutrino mass are shown to be derivable in this 
framework without additional symmetries.  Only complete renormalizable 
theories are considered.  An explicit class of models with $Z_n$ ($n \geq 5$) 
lepton and dark symmetry for Dirac neutrinos is derived, as well as an example 
of $Z_3$ dark symmetry. 
\end{abstract}

 \newpage
 \baselineskip 24pt

\noindent \underline{\it Introduction}~:~
Two outstanding fundamental issues in particle physics and astroparticle 
physics are neutrinos and dark matter.  They have been shown~\cite{m15} 
to be intimately connected in all simple models of dark matter, with 
dark parity $\pi_D$ derivable from lepton parity $\pi_L$ with the factor 
$(-1)^{2j}$ for a particle of spin $j$.

To explore further this connection, it is assumed that lepton number may 
be imposed as a global $U(1)_L$ symmetry in dark-matter extensions of the 
standard model (SM) of quark and lepton interactions.  The extended field 
theory is required to be renormalizable and the $U(1)_L$ symmetry be 
respected by all dimension-four terms in the Lagrangian, whereas the 
soft dimension-three and dimension-two terms are allowed to break $U(1)_L$ 
to $Z_N$.  In particular, the $U(1)_L$ is used to forbid a tree-level 
Majorana or Dirac neutrino mass, whereas its soft breaking will usher in 
a radiative Majorana or Dirac neutrino mass through dark matter, i.e. 
the scotogenic mechanism.  Because of the chosen particle content and 
its original $U(1)_L$ assignments, the resulting theory conserves either 
lepton parity for Majorana neutrinos, or (redefined) lepton number for 
Dirac neutrinos.  At the same time, a dark symmetry also emerges.

\noindent \underline{\it Scotogenic Majorana Neutrino Mass}~:~
Consider an extension of the SM with three Higgs doublets: 
$\Phi = (\phi^+,\phi^0)$, $\eta_1 = (\eta_1^+,\eta_1^0)$, and 
$\eta_2 = (\eta_2^+,\eta_2^0)$, together with three neutral singlet 
left-handed $N_L$ and right-handed $N_R$ fermions.  They are listed 
in Table 1.
\begin{table}[tbh]
\centering
\begin{tabular}{|c|c|c|c|}
\hline
fermion/scalar & $SU(2)$ & $U(1)_Y$ & $U(1)_L$ \\
\hline
$(\nu,e)_L$ & 2 & $-1/2$ & $1$ \\ 
$e_R$ & 1 & $-1$ & 1 \\
$N_L$ & 1 & 0 & $x_1 \neq -1$ \\ 
$N_R$ & 1 & 0 & $x_2 \neq 1$ \\ 
\hline
$\Phi=(\phi^+,\phi^0)$ & 2 & 1/2 & 0 \\ 
$\eta_1=(\eta_1^+,\eta_1^0)$ & 2 & 1/2 & $y \neq 0$ \\ 
$\eta_2=(\eta_2^+,\eta_2^0)$ & 2 & 1/2 & $-y$ \\ 
\hline
\end{tabular}
\caption{Fermion and scalar content of generic model.}
\end{table}
The $U(1)_L$ assignments $x_{1,2}$ are chosen to forbid the tree-level 
couplings $N_L (\nu_L \phi^0 - e_L \phi^+)$ and 
$\bar{N}_R (\nu_L \phi^0 - e_L \phi^+)$.  The choice of $\pm y \neq 0$ 
is to distinguish $\Phi$ from $\eta_{1,2}$ and to allow the quartic 
coupling $(\Phi^\dagger \eta_1)(\Phi^\dagger \eta_2)$ as first proposed 
in Ref.~\cite{mpr13}.  In the original scotogenic model~\cite{m06}, 
$\eta_1=\eta_2$ and is distinguished from $\Phi$ by lepton parity~\cite{m15}.
In the present framework, there are four variations as shown in Fig.~1.
\begin{figure}[htb]
\vspace*{-5cm}
\hspace*{-3cm}
\includegraphics[scale=1.0]{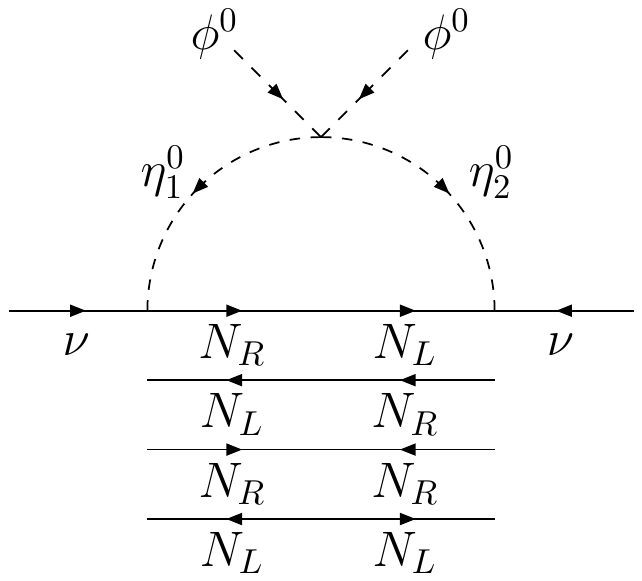}
\vspace*{-19.5cm}
\caption{One-loop diagrams of Majorana neutrino mass.}
\end{figure}

The fermion line has four possible connections. From top to bottom: 
$x_1=y-1, x_2=y+1$; $x_1=-y-1, x_2=-y+1$; $x_2 = y+1 = -y+1$; 
$x_1 = y-1 = -y-1$.  The last two options require $y=0$ which is 
ruled out.  The first two options requires $x_2-x_1=2$. This means that 
the soft term $\bar{N}_L N_R$ must break $U(1)_L$ by two units.  At the 
same time, the soft term $N_L N_L$ breaks it by $2y-2$ or $2y+2$ units, 
whereas the soft term $N_R N_R$ does it by $2y+2$ or $2y-2$ units. 
Furthermore, the soft terms $\Phi^\dagger \eta_{1,2}$ would break $U(1)_L$ 
and allow $\eta^0_{1,2}$ to couple to $\bar{\nu}_L N_R$ through $\phi^0$, 
so they must be forbidden.  To do so, $y$ should be odd, because the 
$\bar{N}_L N_R$ breaking is even (2 units) and these terms will not be 
generated if they are assumed to be absent in the beginning.

If $y=1$, then $x_1=0$ and $x_2=2$.  The resulting symmetry is just 
lepton parity, i.e. $\eta_{1,2}$ are odd and $N_{L,R}$ are even.  
If this symmetry was imposed in the beginning, then $\eta_1=\eta_2$ 
and $N_L = \bar{N}_R$ may be assumed, and the original scotogenic 
model~\cite{m06} is recovered.  However, with soft breaking $U(1)_L$, 
$\eta_1 \neq \eta_2$ and $N_L \neq \bar{N}_R$.  Nevertheless, 
lepton parity $\pi_L$ is still conserved, hence also dark parity 
$\pi_D = \pi_L (-1)^{2j}$ as remarked earlier.  Note that this happens 
for any odd $y$, pointing to the generality of the $U(1)_L$ approach.

In the above example, the dark symmetry is $\pi_D$.  If $U(1)_D$ is 
desired, then it has to be imposed as in Ref.~\cite{mpr13}.  To insist 
on obtaining $U(1)_D$ without imposing it from the beginning, using 
only $U(1)_L$, the following variation may be considered, as shown in 
Table 2.
\begin{table}[tbh]
\centering
\begin{tabular}{|c|c|c|c|}
\hline
fermion/scalar & $SU(2)$ & $U(1)_Y$ & $U(1)_L$ \\
\hline
$(\nu,e)_L$ & 2 & $-1/2$ & $1$ \\ 
$e_R$ & 1 & $-1$ & 1 \\
$E_L$ & 1 & $-1$ & $x_1 \neq -1$ \\ 
$E_R$ & 1 & $-1$ & $x_2 \neq 1$ \\ 
\hline
$\Phi=(\phi^+,\phi^0)$ & 2 & 1/2 & 0 \\ 
$\eta_1=(\eta_1^0,\eta_1^-)$ & 2 & $-1/2$ & $y \neq 0$ \\ 
$\eta_2=(\eta_2^{++},\eta_2^+)$ & 2 & 3/2 & $-y$ \\ 
$\chi^0$ & 1 & 0 & $-y$ \\ 
\hline
\end{tabular}
\caption{Fermion and scalar content of second example.}
\end{table}
The analogous one-loop diagram is shown in Fig.~2.
\begin{figure}[htb]
\vspace*{-5cm}
\hspace*{-3cm}
\includegraphics[scale=1.0]{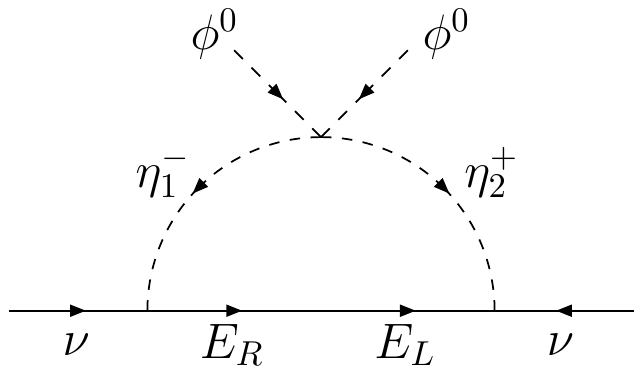}
\vspace*{-21.5cm}
\caption{Scotogenic Majorana neutrino mass with $U(1)_D$.}
\end{figure}
Again $U(1)_L$ is broken by the soft $\bar{E}_L E_R$ mass term by two 
units, but now a dark $U(1)_D$ symmetry remains for 
$\eta_1, E, \eta_2^\dagger$.  On the other hand, $\eta_1^0$ is unsuitable 
as a dark-matter candidate because it couples to the $Z$ boson and thus 
ruled out by underground direct search experiments.  This model (without 
$\chi^0$) was considered previously in Ref.~\cite{aky11}, but dark parity 
was imposed there and lepton number was said to be broken by the dimension-four 
term $(\Phi^\dagger \eta_1)(\Phi^\dagger \eta_2)$, without realizing that its  
correct implemention is softly broken $U(1)_L$ and that a dark $U(1)_D$ 
symmetry remains. Here, the added complex scalar singlet $\chi^0$ is a 
viable dark-matter candidate, providing that the 
$\chi^0 (\eta_1^0 \phi^0 - \eta_1^- \phi^+)$ coupling is suitably 
small.  Again, any odd $y$ works, with $x_1=y-1, x_2=y+1$.  In the special 
case $y=1$, the additional term $\chi^0 \bar{E}_L e_R$ is allowed, with 
further possible interesting phenomenology.  This second example shows the 
power of $U(1)_L$ in combination of the chosen particle content in 
acquiring radiative Majorana neutrino mass together with a dark symmetry.

A third example uses a scalar triplet and a singlet, so that the soft 
bilinear (and trilinear) scalar terms are absent to begin with.  Hence 
there is no constraint on $y$ at this stage.  However, the dark fermions must 
now be doublets and they can form bilinear terms with the SM lepton doublets. 
To prevent their existence, $y$ must again be odd as shown below. 
\begin{table}[tbh]
\centering
\begin{tabular}{|c|c|c|c|}
\hline
fermion/scalar & $SU(2)$ & $U(1)_Y$ & $U(1)_L$ \\
\hline
$(\nu,e)_L$ & 2 & $-1/2$ & $1$ \\ 
$e_R$ & 1 & $-1$ & 1 \\
$(N,E)_L$ & 2 & $-1/2$ & $x_1 \neq 1$ \\ 
$(N,E)_R$ & 2 & $-1/2$ & $x_2 \neq 1$ \\ 
\hline
$\Phi=(\phi^+,\phi^0)$ & 2 & 1/2 & 0 \\ 
$\rho=(\rho^+,\rho^0,\rho^-)$ & 3 & $0$ & $y$ \\ 
$\chi^+$ & 1 & 1 & $-y$ \\ 
\hline
\end{tabular}
\caption{Fermion and scalar content of third example.}
\end{table}
This model was one of the compilations considered in Refs.~\cite{lm13,rzy13}. 
Again their basic assumption was to have dark parity to begin with, in 
which case the scalar triplet $\rho$ may be chosen real.  Here only $U(1)_L$ 
is used, which requires $x_1=y-1,x_2=y+1$ as in the previous two examples. 
The soft term $\bar{E}_L E_R$ again breaks $U(1)_L$ by two units, whereas 
the soft term $\bar{e}_L E_R$ would break it by $y$, so again $y$ must be odd 
to allow the latter to be absent. 
Since $\rho^0$ has no coupling to $Z$, it is a viable dark-matter candidate 
in this case and the dark symmetry $U(1)_D$ emerges as a consequence 
of softly broken $U(1)_L$ together with the chosen particle content.

\noindent \underline{\it Scotogenic Dirac Neutrino Mass}~:
The same idea of using $U(1)_L$ may be applied to Dirac neutrinos. 
Assuming that $U(1)_L$ comes from gauged $B-L$, there have been three recent 
studies~\cite{bccps18,cryz19,jvs19}.  The following analysis shares many of 
their methods and results, but with an important difference.  Whereas they 
consider only dimension-five operators for obtaining radiative Dirac 
neutrino masses, the adopted procedure here is to use dimension-four 
operators with softly broken $U(1)_L$, i.e. the imposition of $U(1)_L$ to 
forbid the tree-level mass, but to allow a radiative mass to appear from 
dimension-two and/or dimension-three terms which break $U(1)_L$, so that 
a dark symmetry emerges as well. 

To have a Dirac neutrino mass, the right-handed singlet neutrino $\nu_R$ 
must exist.  It should pair up with $\nu_L$ through the SM Higgs boson 
$\phi^0$.  Hence it should have $L=1$ under $U(1)_L$.  In that case, 
a tree-level Yukawa coupling is allowed which must however be very small 
to account for the observed neutrino mass limit of 1.1 eV~\cite{katrin19}. 
To forbid this tree-level coupling, a symmetry is routinely applied to 
distinguish $\nu_R$ from the other SM particles, but $L=1$ is retained.  
For a short review, see Ref.~\cite{mp17}.  A generic one-loop diagram is 
depicted in Fig.~3, with its particle content shown in Table 4.
\begin{figure}[htb]
\vspace*{-5cm}
\hspace*{-3cm}
\includegraphics[scale=1.0]{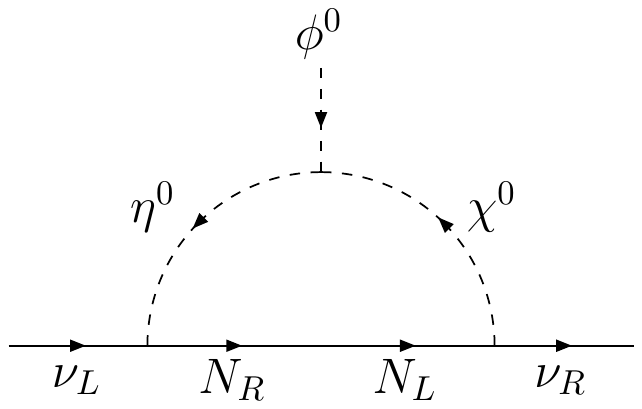}
\vspace*{-21.5cm}
\caption{Scotogenic Dirac neutrino mass.}
\end{figure}
\begin{table}[tbh]
\centering
\begin{tabular}{|c|c|c|c||c|c|c|}
\hline
fermion/scalar & $SU(2)$ & $U(1)_Y$ & $U(1)_L$ & {\large$*$} & $Z_n^L$ & 
$Z_n^D$ \\
\hline
$(\nu,e)_L$ & 2 & $-1/2$ & $1$ & 1 & $\omega$ & 1 \\ 
$e_R$ & 1 & $-1$ & 1 & 1 & $\omega$ & 1 \\
$\nu_R$ & 1 & 0 & $x$ & $-n+1$ & $\omega$ & 1 \\ 
$N_L$ & 1 & $0$ & $y$ & $2-n$ & $\omega^2$ & $\omega$ \\ 
$N_R$ & 1 & $0$ & $y$ & $2-n$ & $\omega^2$ & $\omega$ \\ 
\hline
$\Phi=(\phi^+,\phi^0)$ & 2 & 1/2 & 0 & 0 & 1 & 1 \\ 
$\eta=(\eta^+,\eta^0)$ & 2 & $1/2$ & $y-1$ & $1-n$ & $\omega$ & $\omega$ \\ 
$\chi^0$ & 1 & 0 & $y-x$ & $1$ & $\omega$ & $\omega$ \\ 
\hline
\end{tabular}
\caption{Fermion and scalar content for scotogenic Dirac neutrino mass.}
\end{table}

Here $x \neq 1$ is imposed so that $\nu_R$ does not couple to $\nu_L$ at 
tree level.  To connect them in one loop, the trilinear 
$\bar{\eta}^0 \phi^0 \chi^0$ term must break $U(1)_L$ softly by $x-1$.  
Now $N_L$ and $N_R$ are assumed to have the same $U(1)_L$ charge, i.e. $y$, 
so that $y \neq \pm 1$ and $y \neq \pm x$ are required.  Furthermore, 
$N_L N_L$ or $N_R N_R$ would break $U(1)_L$ by $2y$, $\nu_R \nu_R$ by $2x$, 
$N_R \nu_R$ by $x+y$, $\bar{N}_L \nu_R$ by $x-y$,  and $\chi^0 \chi^0$ 
by $2(y-x)$.  
To allow them to be absent in a complete theory, their $U(1)_L$ charges 
must not be zero, or divisible by the required $U(1)_L$ breaking, i.e. 
$x-1$. To have a solution, $x$ and $y$ must be chosen so that the residual 
symmetry after $U(1)_L$ breaking maintains an effective lepton 
symmetry together with a dark symmetry.

Let $x=-n+1$, then $U(1)_L$ breaks to $Z_n$.  If for example 
$n=3$~\cite{mpr13}, it would be impossible for neutrinos to be Majorana, 
i.e. they must remain Dirac as shown in Ref.~\cite{mr15}. 
The structure of Fig.~3 is well-known~\cite{gs08,fm12}.  It is realized 
conventionally by 3 symmetries: (A) conventional lepton number, where 
$\nu_{L,R},N_{L,R}$ have $L=1$, and $\Phi,\eta,\chi$ have $L=0$, which is 
strictly conserved; (B) dark $Z_2$ symmetry, under which $N_{L,R},\eta,\chi$ 
are odd and others even, which is strictly conserved; and (C) an 
{\it ad hoc} $Z_2$ symmetry under which $\nu_R,\chi$ are odd and all 
others even, which is softly broken by the $\eta^\dagger \Phi \chi$ term.  
In previous applications, 
$\chi$ is assumed to be a real neutral scalar singlet for simplicity.  Here 
it is crucial that it is complex to carry the nonzero $U(1)_L$ charge $y-x$. 

In Table 4, in the column denoted by $*$, the $U(1)_L$ charges are 
chosen explicitly. The terms $\bar{N}_L \nu_R$ and $\chi^0 \chi^0$ have 
$U(1)_L$ charges $-1$ and $2$.  They are not zero or divisible by $n \geq 3$. 
The terms $\nu_R \nu_R$, $N_R \nu_R$ and $N_R N_R$ should also be absent, 
their $U(1)_L$ charges divided by $n$ are $(2/n)-2$, $(3/n)-2$ and 
$(4/n)-2$, hence $n=3,4$ are ruled out.  All higher values of $n$ are 
acceptable.  The resulting theory allows two related symmetries: 
(I) $Z_n^L$ lepton symmetry 
under which $\nu_{L,R}, e_{L,R}, \eta, \chi \sim \omega$ and 
$N_{L,R} \sim \omega^2$, where $\omega^n=1$; (II) $Z_n^D$ dark symmetry, 
derivable from lepton symmetry by multiplying the latter by $\omega^{-2j}$ 
where $j$ is the particle's spin.  As a result, $\nu_{L,R}, e_{L,R} \sim 1$ 
and $N_{L,R}, \eta, \chi \sim \omega$.  This is the Dirac generalization of 
the Majorana case of the derivation of dark parity $\pi_D$ from lepton 
parity $\pi_L$ first pointed out in Ref.~\cite{m15}.

In a renormalizable theory, $Z_n$ symmetry is not simply realizable for 
large $n$ because the Lagrangian admits only terms of dimension four or 
less.  In the above example for $n \geq 5$, the Lagrangian cannot admit the 
term $(\chi^0)^n$, hence the true symmetry of the theory is a redefined 
$U(1)_L$, where $\nu_{L,R}, e_{L,R}, \eta, \chi \sim 1$ and $N_{L,R} \sim 2$. 
The dark symmetry is then $U(1)_D$ where it is derived from $U(1)_L$ by 
subtracting from the latter $2j$ where $j$ is the particle's spin, 
i.e. $\nu_{L,R}, e_{L,R} \sim 0$ and $N_{L,R}, \eta, \chi \sim 1$. 
This shows that scotogenic Dirac neutrino mass is derivable from 
softly broken $U(1)_L$ alone with emergent $U(1)$ lepton and dark 
symmetries.

If $Z_n$ symmetry is desired, the scalar sector must be extended.  For 
example, if $n=5$, in addition to $\chi \sim \omega$, another scalar 
$\sigma \sim \omega^3$ may be added.  The coexisting terms $\chi^3 \sigma^*$ 
and $\chi^2 \sigma$ would then enforce $Z^D_5$ as a dark symmetry, but the 
lepton symmetry would become global $U(1)$ as conventionally defined, i.e. 
$L=1$ for $\nu_{L,R}$ and $N_{L,R}$.  To enforce $Z^L_5$ as a lepton symmetry, 
a third scalar $\kappa$ may be added with $U(1)_L$ charge = 7.  In that 
case, the terms $\chi \sigma^2 \kappa^*$ and $\kappa N_R \nu_R$ would 
allow $\nu_{L,R}, \chi, \eta \sim \omega$ and $N_{L,R}, \kappa \sim \omega^2$ 
under $Z^L_5$ with $\omega^5=1$.

Since $\chi^0$ mixes with $\eta^0$, the neutral scalar of the dark sector 
of this model requires this mixing to be very small and the lighter 
eigenstate to be mostly $\chi^0$ for it to be a viable dark-matter 
candidate.  Alternatively the lightest $N$ may also be chosen as dark 
matter.  For details, see Ref.~\cite{fm12}. 

A possible variation of the model is to add a neutral scalar singlet 
$\zeta$ with $U(1)_L$ charge $n$, and require that $U(1)_L$ be spontaneously 
broken only.  In that case, the term $\eta^\dagger \Phi \chi$ is replaced 
with $\zeta^* \eta^\dagger \Phi \chi$.  Neutrinos obtain radiative Dirac 
masses as before, with new emergent $U(1)$ lepton and dark symmetries, 
but now a massless Goldstone boson appears. It is the analog of the 
majoron which comes from breaking $U(1)_L$ spontaneously to $Z_2$ and  
is applicable to Majorana neutrinos, whereas here it is the massless 
diracon~\cite{bv16} which comes from breaking $U(1)_L$ spontaneously 
to $Z_n$ $(n \geq 5)$ and is applicable to Dirac neutrinos.

There is a further use of $\zeta$, if it is allowed to couple anomalously 
to a pair of exotic quarks (color fermion triplets) or a color fermion 
octet~\cite{dm00,mot17}.  Then this diracon becomes a QCD (quantum 
chromodynamics) axion and 
$U(1)_L$ is an extended version of Peccei-Quinn symmetry, as proposed 
many years ago~\cite{s87,m01} for Majorana neutrinos, and very recently 
also for Dirac neutrinos~\cite{prsv19,b19}.  In these scenarios, dark 
matter consists of both the axion and a WIMP (weakly interacting massive 
particle)~\cite{dmt14}.

In Fig.~3, the fermion singlets $N_{L,R}$ may be replaced with the doublets 
$(E^0,E^-)_{L,R}$ as shown in Table 5.  
\begin{table}[tbh]
\centering
\begin{tabular}{|c|c|c|c||c|c|c|}
\hline
fermion/scalar & $SU(2)$ & $U(1)_Y$ & $U(1)_L$ & {\large$**$} & $L$ & 
$Z_3^D$ \\
\hline
$(\nu,e)_L$ & 2 & $-1/2$ & $1$ & 1 & $1$ & 1 \\ 
$e_R$ & 1 & $-1$ & 1 & 1 & $1$ & 1 \\
$\nu_R$ & 1 & 0 & $x$ & $-2$ & $1$ & 1 \\ 
$(E^0,E^-)_L$ & 2 & $-1/2$ & $y$ & $2$ & $1$ & $\omega$ \\ 
$(E^0,E^-)_R$ & 2 & $-1/2$ & $y$ & $2$ & $1$ & $\omega$ \\ 
\hline
$\Phi=(\phi^+,\phi^0)$ & 2 & 1/2 & 0 & 0 & 0 & 1 \\ 
$\eta=(\eta^+,\eta^0)$ & 2 & $1/2$ & $x-y$ & $-4$ & $0$ & 
$\omega^{-1}$ \\ 
$\chi^0$ & 1 & 0 & $y-1$ & $1$ & $0$ & $\omega$ \\ 
\hline
\end{tabular}
\caption{Fermion and scalar content for scotogenic $Z_3$ Dirac neutrino mass.}
\end{table}
This construction eliminates the existence of many fermion bilinears 
except $\nu_R \nu_R$ and $\bar{\nu}_L E^0_R + e^+_L E^-_R$.  Hence only 
$2x$ and $y-1$ must not be zero or divisible by $x-1$.  Also $y \neq x$ 
is required.  As a result, it is possible to have $Z_3$ dark 
symmetry, i.e. $x=-2$ and $y=2$, as shown in the column denoted by $**$.  
The analogous one-loop diagram for scotogenic Dirac neutrino mass is shown 
in Fig.~4.   
\begin{figure}[htb]
\vspace*{-5cm}
\hspace*{-3cm}
\includegraphics[scale=1.0]{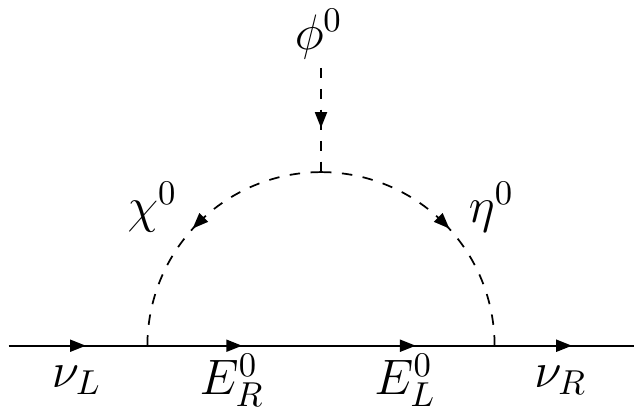}
\vspace*{-21.5cm}
\caption{Scotogenic Dirac neutrino mass with $Z_3$ dark symmetry.}
\end{figure}
The $U(1)_L$ symmetry is broken to $Z_3$ by the soft trilinear 
scalar terms $\Phi^\dagger \eta \chi$ and $\chi^3$.  However, the 
dimension-four term $\chi^0 \nu_R \nu_R$ which is allowed by $Z_3$ 
is not allowed by the original $U(1)_L$.  Hence the breaking does not 
affect the fermions of this model and the conventional assignment of 
$L=1$ may be applied to $\nu_{L,R}, E^0_{L,R}$ with $L=0$ for all the 
scalars.  The $Z^D_3$ dark symmetry emerges as before.

In this example, the $U(1)_L$ global symmetry is anomalous.  To make it 
anomaly-free so that it can be promoted to a gauge symmetry, the three 
copies of $\nu_R$ with charge $-2$ should be changed to $1$, as in the 
conventional assignments for gauge $B-L$.  The difference is then 
$3(1)-3(-2)=9$ for the sum of $U(1)_L$ charges and $3(1)-3(-8)=27$ 
for the sum of the cubes of the charges.  A complete renormalizable 
anomaly-free gauge $U(1)_L$ theory is then possible with the following 
additional particle content.

Singlet right-handed fermions $\psi_{2,3,4}$ are added with $U(1)_L$ 
charges $-2,3,-4$ respectively.  Let there be 3 copies each of $\psi_{2,4}$ 
and 9 copies of $\psi_3$.  Then $3(-2-4)+9(3)=9$ and 
$3(-8-64)+9(27)=27$, satisfying the requirement for the theory to be 
anomaly-free.  To break the gauge $U(1)_L$ symmetry to $Z_3$, the scalar 
singlets $\zeta_{3,6}$ with charges $3,6$ are used, so that the terms  
$\chi^3 \zeta_3^*$, $\zeta_3^2 \zeta_6^*$, $\psi_3 \psi_3 \zeta_6^*$, 
and $\psi_2 \psi_4 \zeta_6$ are allowed in the complete Lagrangian, 
ensuring that all new fermions acquire nonzero masses.  The new fermions 
$\psi_{2,3,4}$ have $L=1,0,-1$ and transform trivially under $Z_3^D$.
In addition $\psi_3$ has its own accidental (or predestined) $Z_2$ symmetry 
from the chosen particle content of the theory and the imposed $U(1)_L$ 
symmetry.  The Dirac neutrino mass 
matrix is now $6 \times 6$ with 3 tree-level masses and 3 one-loop masses, 
the latter linking only to the left-handed SM neutrinos.  However, there 
could be mixing between the two sectors which may be a source of 
nonunitarity of the observed $3 \times 3$ neutrino mixing matrix.

\noindent \underline{\it Conclusion}~:~
The intrinsic connection between lepton number and dark symmetry has 
been demonstrated with three examples in the case of Majorana neutrinos 
where $U(1)_L$ is broken softly to $Z_2$ lepton parity $\pi_L$.  
In the first example, $Z_2$ dark parity $\pi_D = \pi_L (-1)^{2j}$ 
emerges and scotogenic Majorana neutrino mass is obtained.  In the 
second and third examples, by choosing different particles in the 
dark sector, a dark $U(1)_D$ symmetry is maintained.

Using the same connection, two examples of scotogenic Dirac neutrino mass 
have also been described, one with emergent $Z_n$ lepton and dark symmetry 
for $n \geq 5$.  However, without enlarging the necessary scalar sector, 
the requirement of renormalizability of these models implies that the 
true symmetry of the Lagrangian is a redefined $U(1)_L$ such that a dark 
$U(1)_D$ is obtained by subtracting the assigned lepton number of a 
particle by $2j$ where $j$ is the particle's spin.

The other example chooses a different set of dark fermions so that 
$Z_3$ dark symmetry emerges which is maintained explicitly by the 
renormalizable Lagrangian of the model.  It also sustains a conserved 
lepton number in the conventional way.

These two examples generalize the case of $Z_2$ lepton and dark 
parity~\cite{m15} for Majorana neutrinos to Dirac neutrinos.

\noindent \underline{\it Acknowledgement}~:~
This work was supported in part by the U.~S.~Department of Energy Grant 
No. DE-SC0008541.

\bibliographystyle{unsrt}

\end{document}